\documentclass[aapm,graphicx]{revtex4-1}
\draft 

\usepackage{amssymb}
\usepackage{graphicx}
\usepackage{multirow}
\usepackage{color,soul}

\graphicspath{{./}}

\begin{document}

\title{On the use of machine learning methods for mPSD calibration in HDR brachytherapy} 

\author{Haydee M. Linares Rosales$^{1,2}$}
\author{Gabriel Couture$^{1,2}$}
\author{Louis Archambault$^{1,2}$}
\author{Sam Beddar$^{3, 4}$}
\author{Philippe Despr\'es$^{1,2}$}
\author{Luc Beaulieu$^{1,2}$}

\affiliation{$^{1}$D\'epartement de physique, de g\'enie physique et d'optique et Centre de recherche sur le cancer, Universit\'e Laval, Qu\'ebec, Canada.}
\affiliation{$^{2}$D\'epartement de radio-oncologie et Axe Oncologie du CRCHU de Qu\'ebec, CHU de Qu\'ebec - Universit\'e Laval, QC, Canada.}
\affiliation{$^{3}$Department of Radiation Physics, The University of Texas MD Anderson Cancer Center, Houston,TX, United States.}
\affiliation{$^{4}$The University of Texas MD Anderson UTHealth Graduate School of Biomedical Sciences, Houston, TX, United States.}

\email[Corresponding author: Luc Beaulieu, ]{Luc.Beaulieu@phy.ulaval.ca}
\date{\today}

\begin{abstract}
\scriptsize{\textbf{Purpose:} We sought to evaluate the feasibility of using machine learning (ML) algorithms for multipoint plastic scintillator detector (mPSD) calibration in high-dose-rate (HDR) brachytherapy.\\

\textbf{Methods}: Dose measurements were conducted under HDR brachytherapy conditions. The dosimetry system consisted of an optimized 1-mm-core mPSD and a compact assembly of photomultiplier tubes coupled with dichroic mirrors and filters. An $^{192}Ir$ source was remotely controlled and sent to various positions in a homemade PMMA holder, ensuring 0.1-mm positional accuracy. Dose measurements covering a range of 0.5 to 12 cm of source displacement were carried out according to TG-43 U1 recommendations. Individual scintillator doses were decoupled using a linear regression model, a random forest estimator, and artificial neural network algorithms.  The dose predicted by the TG-43U1 formalism was used as the reference for system calibration and ML algorithm training. The performance of the different algorithms was evaluated using different sample sizes and distances to the source for the mPSD system calibration. \\

We found that the calibration conditions influenced the accuracy in predicting the measured dose. The decoupling methods' deviations from the expected TG-43 U1 dose generally remained below 20 \%. However, the dose prediction with the three algorithms was accurate to within 7 \% relative to the dose predicted by the TG-43 U1 formalism when measurements doses were performed in the same range of distances used for calibration. In such cases, the predictions with random forest exhibited minimal deviations ($<$ 2 \%). However, the performance random forest was compromised when the predictions were done beyond the range of distances used for calibration. Because the linear regression algorithm can extrapolate the data, the dose prediction by the linear regression was less influenced by the calibration conditions than random forest. The linear regression algorithm's behavior along the distances to the source was smoother than those for the random forest and neural network algorithms, but the observed deviations were more significant than those for the neural network and random forest algorithms. The number of available measurements for training purposes influenced the random forest and neural network models the most. Their accuracy tended to converge toward deviation values close to 1\% from a number of dwell positions greater than 100. 

\textbf{Conclusions:} In performing HDR brachytherapy dose measurements with an optimized mPSD system, ML algorithms are good alternatives for precise dose reporting and treatment assessment during this kind of cancer treatment.

}

\end{abstract}


\maketitle 
 
\section{Introduction}

    Several studies \cite{Fonseca-2017, Cartwright-2010, Hardcastle-MOSkin-2010, Kertzscher-2011, Seymour-2011, Kertzscher-2014, Kertzscher-Inorganic-Scint-2017, Kertzscher-Inorganic-scint-2019, Johansen-2018, Sethi-Doppler-US-2018} have focused on developing detectors and methods for real-time source monitoring in brachytherapy. The benefits and limitations of plastic scintillation detectors (PSDs) in a single or multi-point configuration (multipoint PSD [mPSD]) have been well documented \cite{Therriault-Temp-method-2015, Boivin-2016, Lambert-Cerenkov-2008, Beddar-Cerenkov-1992, Archambault-2006, Wootton-Temperature-2013, Guillot-toward-2010, Beddar-water-equivalent-1992-1, Beddar-water-equivalent-1992-2, Beaulieu-Scint-Status-2013, Beaulieu-Review-2016, Beddar-temp-2012,  Linares-2019}. In a recent study, Linares Rosales \textit{et al.} \cite{Linares-2019} demonstrated that with proper optimization of the optical chain in combination with the mathematical formalism proposed in \cite{Archambault-MathForm-2012}, the use mPSD is suitable for real-time source tracking in high dose rate (HDR) brachytherapy. 
    
    The goal of the present prospective study was to evaluate the feasibility of machine learning (ML) algorithms for mPSD calibration in HDR brachytherapy based on prior mapping of the response of the mPSD to different radiation source positions.  We compared three methods of scintillator dose decoupling: (1) a linear regression model \cite{Archambault-MathForm-2012, Linares-2019}; (2) a random forest estimator, and; (3) artificial neural network (ANN) algorithms. To the best of our knowledge, no previous reported studies have examined the use of ML algorithms for scintillator detector calibration in HDR brachytherapy.

\section{Materials and Methods}

    \subsection{HDR brachytherapy dose measurements}
    
    \subsubsection{Experimental setup}

    Dose measurements were performed in HDR brachytherapy, under full TG43 conditions \cite{TG-43-Update, Perez-2012-HEBD}. The dosimetry system used for the measurements compriseds three3 parts: 1) a fully optimized and characterized multi-point plastic scintillator dosimeter (1- mm- core mPSD; using BCF-60, BCF-12, and BCF-10 scintillators) \cite{Linares-2019}, 2) a light detection system, and 3) a Python-based graphical user interface for system management and signal processing. The light detection system was composed of a compact assembly of photomultiplier tubes coupled with dichroic mirrors and filters for high-sensitivity scintillation light collection \cite{Linares-2019}. The photomultiplier tubes were independently controlled and read simultaneously at a rate of 100 kHz using a USB-6216 M Series multifunction data acquisition board (National Instruments, Austin, TX).
    
    \begin{figure}[ht]
    \centering
    \begin{tabular}{c}
    \includegraphics[trim = 1mm 1mm 1mm 1mm, clip, scale=0.43]{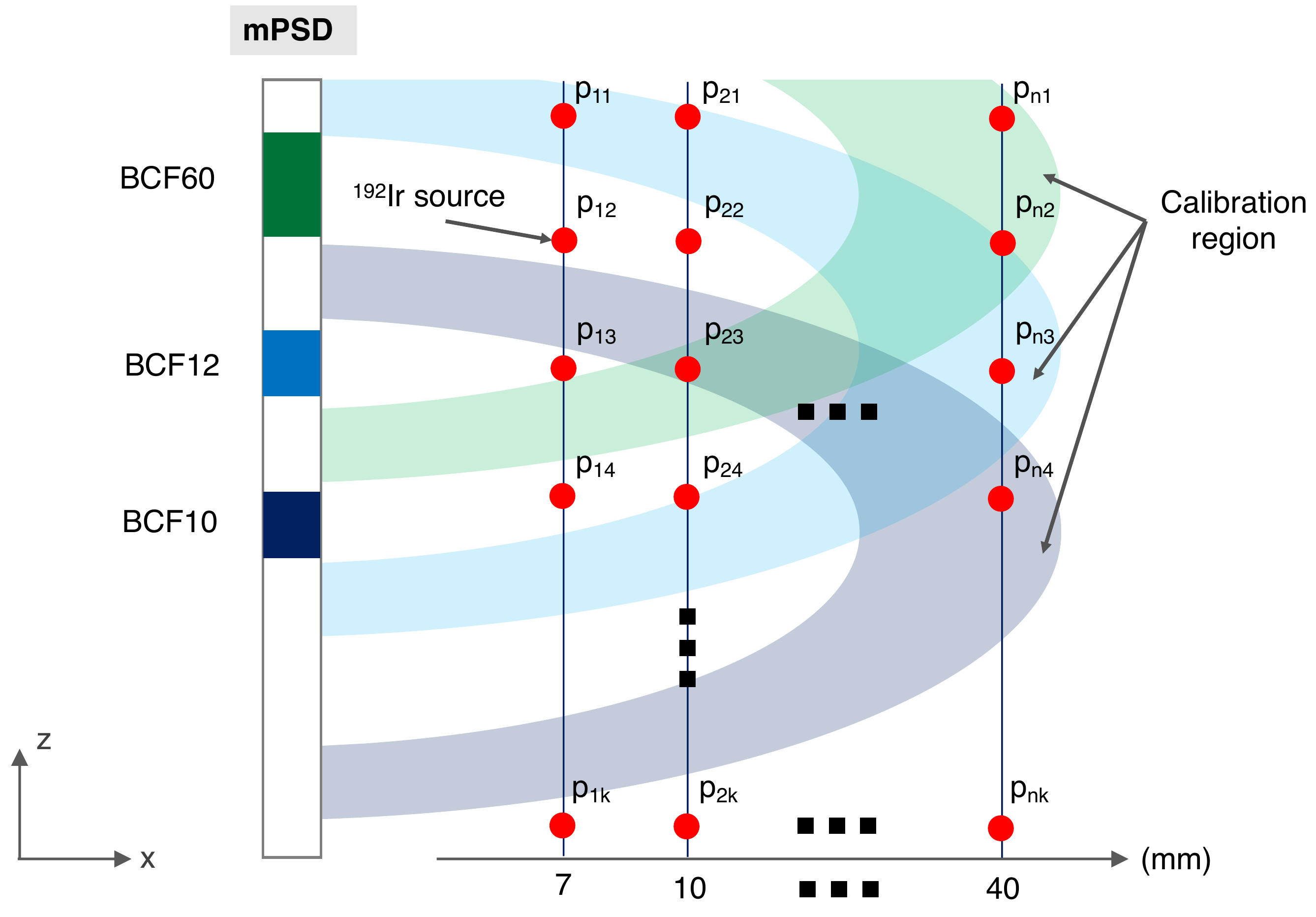}
    \end{tabular}
    \caption{\label{MM_fig_meas_schematic} Schematic of the source positions and nomenclature used for data training with ML algorithms during HDR brachytherapy dose measurements. The circular shaded regions around the scintillator centers represent the regions of radial distances used during the system calibration.}
    \end{figure}
    
    Figure \ref{MM_fig_meas_schematic} shows a schematic of the source positions with respect to the scintillators and the nomenclature used in this work. Each source dwell position was labeled as $p_{nk}$, where $n$ represents the catheter number and $k$ represents the point of measurement inside the catheter. Sixteen catheters were positioned in a PMMA holder \cite{Therriault-mPSD-Brachy-2013, Linares-2019}, covering a 40-mm range of source displacement in the x direction. The source dwelled at 121 locations inside each catheter with 1-mm steps for a total of 1936 independent dwell positions. A source dwell time of 1 s was planned for each position. Thus, for a single source dwell position, $1 \times 10^{5}$ samples were collected. The circular-shaped shaded region around each scintillator center represents the region of radial distances considered during the system calibration (see section \ref{MM_section_influence_calib} for details).

\subsubsection{Dose calculation model}

     According to a formalism for a three-point mPSD described previously \cite{Linares-2019} the signal produced by the source at position $p_{nk}$, can be described by the voltage function $V_{nk} = f([V^{1}_{nk}, V^{2}_{nk}, V^{3}_{nk}, V^{4}_{nk}])$, which is a combination of the four voltages ($V^{1}_{nk}, V^{2}_{nk}, V^{3}_{nk}, V^{4}_{nk}$) collected at each measurement channel of the mPSD system \cite{Linares-2019}. Dose calculation with mPSDs is based on the assumption that the recorded signal results from the linear superposition of spectra; no self-absorption interactions among the scintillators in the mPSD are considered \cite{Archambault-MathForm-2012}.
     
     Individual scintillator doses were decoupled using: 1) a linear regression model, 2) a random forest estimator and, 3) an artificial neural network (ANN) algorithm.  The linear regression model used for dose estimation is a hyperspectral approach previously for a three-point mPSD  \cite{Archambault-MathForm-2012, Linares-2019}. A random forest is a meta-estimator that fits a number of classifying decision trees on various sub-samples of the data set and uses averaging to improve the predictive accuracy and control over-fitting. A regression tree is an efficient way of mapping a complex input space to continuous output parameters. Highly nonlinear mappings are handled by splitting the original problem into a set of smaller problems that can be addressed with simple predictors \cite{2013-Criminisi-RegressionForest}. The linear regression and random forest dose calculations were done using the \textit{scikit-learn} Python library \cite{scikit-learn-Python-RF}. Also, an ANN is a family of algorithms involving several processing layers that learn representations of data. A fully connected network was used when doing the dose calculations with the ANN. The ANN was implemented with the \textit{version 2.2.4} of the \textit{Keras} framework \cite{keras-ANN}, relying on the \textit{TensorFlow} backend. 
     
     Table \ref{table_ML_hyper_parameters} summarizes the hyperparameters used to train each dose prediction model. The hyperparameters that are not specified were set to default levels. The random forest's hyperparameters were optimized using the grid search optimization method \cite{Bergstra-hyper-param-ML-2012}. The ANN architecture was optimized using a manual search algorithm \cite{Bergstra-hyper-param-ML-2012}. The input layer in the ANN had four neurons corresponding to the number of measurement channels in the mPSD system, whereas the output layer had three neurons corresponding to the received dose at each scintillator at a given source dwell position ($p_{nk}$). The number of neurons per layer used was 96, except in the study cases 2F-I in table \ref{table_MM_ML_parameters}. In those cases, to avoid overfitting of the dose, the number of neurons was set equal to the number of dwells used as training data. The early stopping method was used to monitor the loss function during the training phase. When the accuracy is not improved over 100 epochs, the training phase is stopped and outputs the best combination of weights. 
 
      \begin{table}[h]
      \caption{\label{table_ML_hyper_parameters}  Hyper-parameters used for training of the dose prediction models \cite{scikit-learn-Python-RF, keras-ANN}.}
      \setlength{\tabcolsep}{10pt}
          \centering
          \begin{tabular}{ccc}
            \hline
            \hline
            \textbf{Linear Regression}        & \textbf{Random Forest}        & \textbf{Neural Network} \\
            \hline
             \multicolumn{1}{l}{\texttt{fit intercept =} False}    & \multicolumn{1}{l}{\texttt{number of estimators =} 300}        & \multicolumn{1}{l}{\texttt{hidden layers = } 2} \\
             \multicolumn{1}{l}{\texttt{normalize =} False}        & \multicolumn{1}{l}{\texttt{maximum depth =} 80}                & \multicolumn{1}{l}{\texttt{neurons per layer =} 96$^{*}$}  \\
             \multicolumn{1}{l}{}            & \multicolumn{1}{l}{\texttt{criterion =} mean square error}     & \multicolumn{1}{l}{\texttt{Adam optimizer's learning rate =} 0.005} \\
             \multicolumn{1}{l}{}    & \multicolumn{1}{l}{\texttt{minimum sample to split =} 2}       & \multicolumn{1}{l}{\texttt{batch size =} 16} \\
                                                                   &                                                                & \multicolumn{1}{l}{\texttt{number of epochs =} 1000} \\
                                                                   &                                                                & \multicolumn{1}{l}{\texttt{loss function =} mean absolute error} \\
                                                                   &                                                                & \multicolumn{1}{l}{\texttt{activation function of hidden layer =} ReLu} \\
                                                                   &                                                                & \multicolumn{1}{l}{\texttt{activation function of output layer =} Linear} \\
                                                                   &                                                                & \multicolumn{1}{l}{\texttt{weight initialization =} normal distribution} \\
                                                                   &                                                                & \multicolumn{1}{l}{\texttt{validation split ratio =} 0.2} \\
                                                                   &                                                                & \multicolumn{1}{l}{\texttt{early stopping patience =} 100} \\
             \hline
              \multicolumn{3}{l}{\scriptsize{$*$ The number of neurons per layer used was 96 all over the paper, except in the study cases 2F-I in table \ref{table_MM_ML_parameters}. In those cases,}}\\
              \multicolumn{3}{l}{\scriptsize{to avoid over-fitting, the number of neurons was set equal to the number of dwells used as training data.}}
              
          \end{tabular}
      \end{table}
      
      The dose predicted by the TG-43 U1 formalism \cite{TG-43-Update} was used as the reference for the system calibration and training. The finite size of each scintillator was accounted for during calculations; the TG-43 U1 dose values were integrated over each scintillator's sensitive volume.

 \subsection{Evaluation of the influence of calibration on the dose prediction model}
 
 \label{MM_section_influence_calib}
 
     The choice of calibration conditions influences the accuracy and noise of the measured dose \cite{Andersen-time-resolved-2009, Linares-2019}. The measurement database was composed of the mPSD signals collected at 1936 dwell positions. Our first step was to randomly split the measurements' database into a training and testing data sets. The training data set was fixed as the 70 \% (1355 dwell positions) of the measurements recorded at all the distances to the source (0-100 mm range). The testing data set was the remaining 30 \% of the measurements (580 dwell positions), and was used to evaluate the model’s performance in dose prediction. The performance of different algorithms was compared when using different sample sizes and distances to the source for the mPSD system calibration. Thus, region mappings were radially created around each scintillator center according to the distance to the source (see figure \ref{MM_fig_meas_schematic} for visual reference). Two criteria were used to assess each model's performance as a function of the distance to the source used for calibration. The first criterion, referred to as the \textit{inside region}, evaluated the model performance in the same region in which the calibration was performed. The second criterion, referred to as the \textit{outside region}, evaluated the performance at distances outside the calibration region. 
     
     Table \ref{table_MM_ML_parameters} summarizes the parameters used for defining the training and testing data sets in each case studied. The main goal of the study case 1 was to evaluate the impact of the distance to the source selection. With study case 2, the impact of the sampling size on the algorithms' dose prediction was assessed. The sampling size in Table \ref{table_MM_ML_parameters} refers to the number of measurement points (i.e., dwell positions) considered for training and/or testing. The second and third columns in Table \ref{table_MM_ML_parameters} list the ranges of distances considered during the training and testing processes. For the training data, those in the training data set falling in the desired region of interest for calibration were selected. Next, a percentage or number of dwell positions to be used as training data was randomly selected. Columns four to seven in Table \ref{table_MM_ML_parameters} show information related to the sampling size used to train and test the algorithms. Column four lists the percentages of the training data used (relative to the available measurement database). Column five lists the numbers of dwell positions used for training purposes. For study case 1, 677 dwell positions were randomly selected from the training data set. Because the primary goal for study case 1 was to study the impact of the distance to the source and the available measurement data fluctuated according to each range of distances to the source, the training data were fixed to 677 dwell positions. For study case 2, different sampling sizes were randomly selected from the training data set over the entire range of distances to the source explored. Because the training data are randomly selected, when a small number of dwell positions is involved in the calibration, the accuracy of the dose prediction can be compromised by non optimal selection of the dwell position for mPSD calibration. Because of the aforementioned, for study cases 2F-I, the training data were randomly selected 150 times, and the system training was performed as well.

     \begin{table}[ht]
    \caption{\label{table_MM_ML_parameters}  Data selection parameters for mPSD calibration with ML algorithms.}
    \setlength{\tabcolsep}{8pt}
    \centering
    
    \vspace{0.2cm}
    	\begin{tabular}{cccccccccccc}
    	\hline
    	\hline
                              & \multicolumn{5}{c}{\textbf{Distance to the source }}                            & & \multicolumn{5}{c}{\textbf{Sampling size}} \\
        \textbf{Study case ID}   & \multicolumn{2}{c}{\textbf{Training}}  & & \multicolumn{2}{c}{\textbf{Testing}} & & \multicolumn{2}{c}{\textbf{Training}} & &\multicolumn{2}{c}{\textbf{Testing}} \\ \cline{2-3}\cline{5-6}\cline{8-9}\cline{11-12}
    	                      & \multicolumn{2}{c}{\textbf{(mm)}}      & & \multicolumn{2}{c}{\textbf{(mm)}}    & & \textbf{(\%)}  & \textbf{(\# of dwell)}    & &\textbf{(\%)} & \textbf{(\# of dwell)}\\
    	\hline
    	\multirow{5}{*}{1}
        \hspace{3mm}       A  &  \multicolumn{2}{c}{0 - 50}                         & & \multicolumn{2}{c}{\multirow{5}{*}{0 - 100}}  && \multirow{5}{*}{35} & \multirow{5}{*}{677} & &\multirow{5}{*}{30} & \multirow{5}{*}{580}\\
        \hspace{5mm}       B  &  \multicolumn{2}{c}{25 - 75}                        & & &                       &&  &                    & &                     \\
        \hspace{5mm}       C  &  \multicolumn{2}{c}{50 - 100}                       & & &                       &&  &                    & &                     \\
        \hspace{5mm}       D  &  \multicolumn{2}{c}{0 - 25 \& 75 - 100}             & & &                       &&  &                    & &                     \\
        \hspace{5mm}       E  &  \multicolumn{2}{c}{0 - 100}                        & & &                       &&  &                    & &                     \\
        \hline
        \multirow{7}{*}{2} 
        \hspace{3mm}       F  &  \multicolumn{2}{c}{\multirow{7}{*}{0 - 100}}       & & \multicolumn{2}{c}{\multirow{7}{*}{0 - 100}}     & & 0.2               & 4     & &  \multirow{7}{*}{30} & \multirow{7}{*}{580} \\
        \hspace{5mm}       G  &                                 & &                               & & & & 0.5               & 10    & &                      &                      \\
        \hspace{5mm}       H  &                                 & &                               & & & & 5                 & 94    & &                      &                      \\
        \hspace{5mm}       I  &                                 & &                               & & & & 10                & 189   & &                      &                      \\
        \hspace{5mm}       J  &                                 & &                               & & & & 30                & 580   & &                      &                      \\
        \hspace{5mm}       K  &                                 & &                               & & & & 50                & 944   & &                      &                      \\
        \hspace{5mm}       L  &                                 & &                               & & & & 70                & 1360  & &                      &                      \\
        \hline
    	\end{tabular}
    \end{table}
     
    \section{Results}
    
    \subsection{The influence of the calibration on the dose prediction model}
    
    \subsubsection{Distance to the source}
    
\begin{figure}[ht]
    \centering
    \begin{tabular}{c}
    \includegraphics[trim = 1mm 1mm 1mm 1mm, clip, scale=0.40]{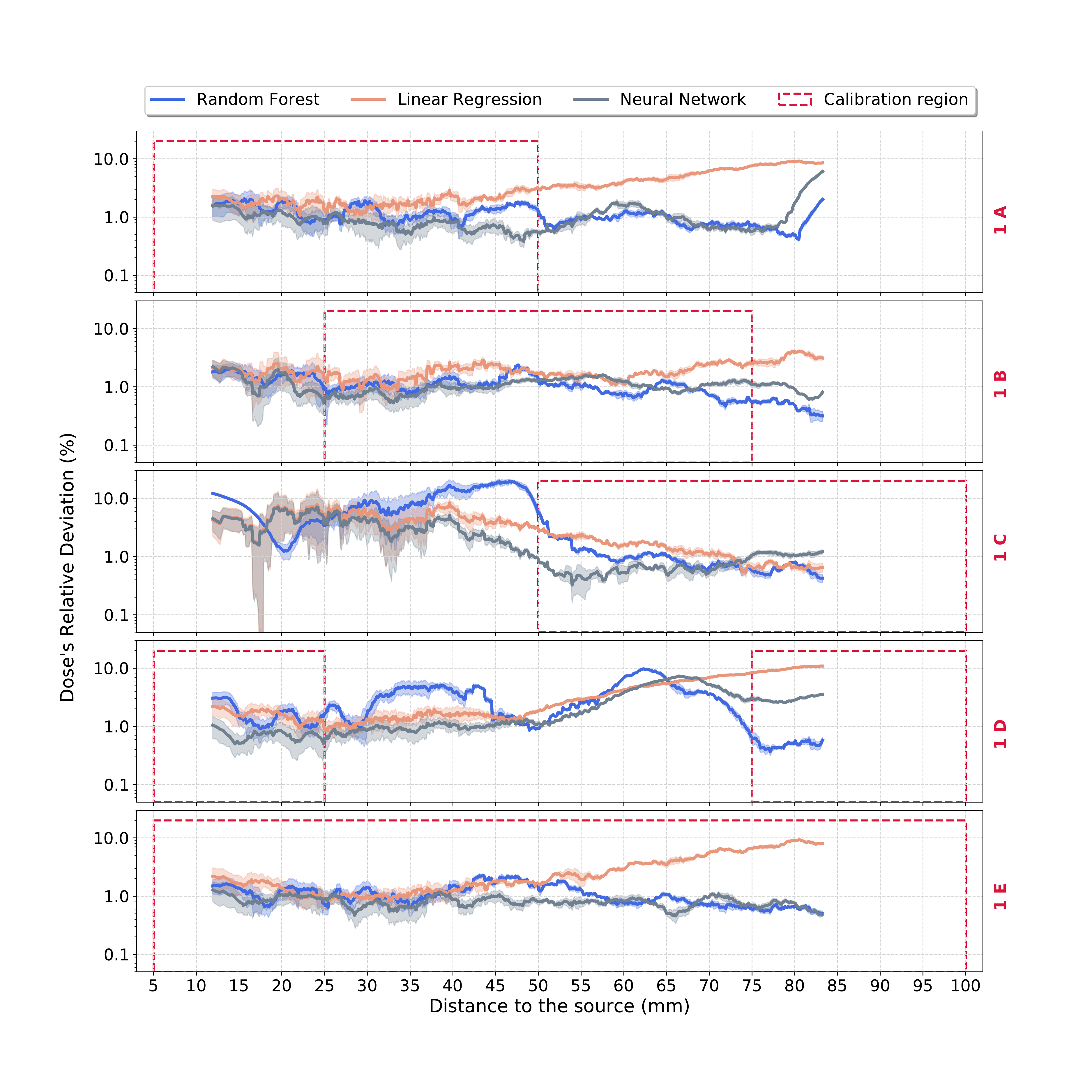}
    \end{tabular}
    \caption{\label{R_influence_calibration} Influence of the calibration region on the dose prediction model.}
    \end{figure}

    Figure \ref{R_influence_calibration} shows the relative deviations (i.e., [measured - expected]/expected) from the expected TG-43 U1 \cite{TG-43-Update} dose obtained with the random forest, linear regression, and ANN algorithms for the sensors in the mPSD. For visualization purposes, the figure only illustrates the results for the BCF-10 scintillator because the results for the BCF-12 and BCF-60 scintillators demonstrated the same behavior. Each solid line in the figure represents the median value of the distribution of deviations at each dwell position. The shaded regions contouring the models' median values represent the standard deviations. The dashed rectangles highlight the regions used for mPSD calibration. The correspondence of the regions to the study cases in Table \ref{table_MM_ML_parameters} is indicated on the right side. The columns in Figure \ref{R_influence_calibration} represent the results for each scintillator in the mPSD.
    
    \subsubsection{Sample size}
    
\begin{figure}[ht]
    \centering
    \begin{tabular}{c}
    \includegraphics[trim = 1mm 1mm 1mm 1mm, clip, scale=0.40]{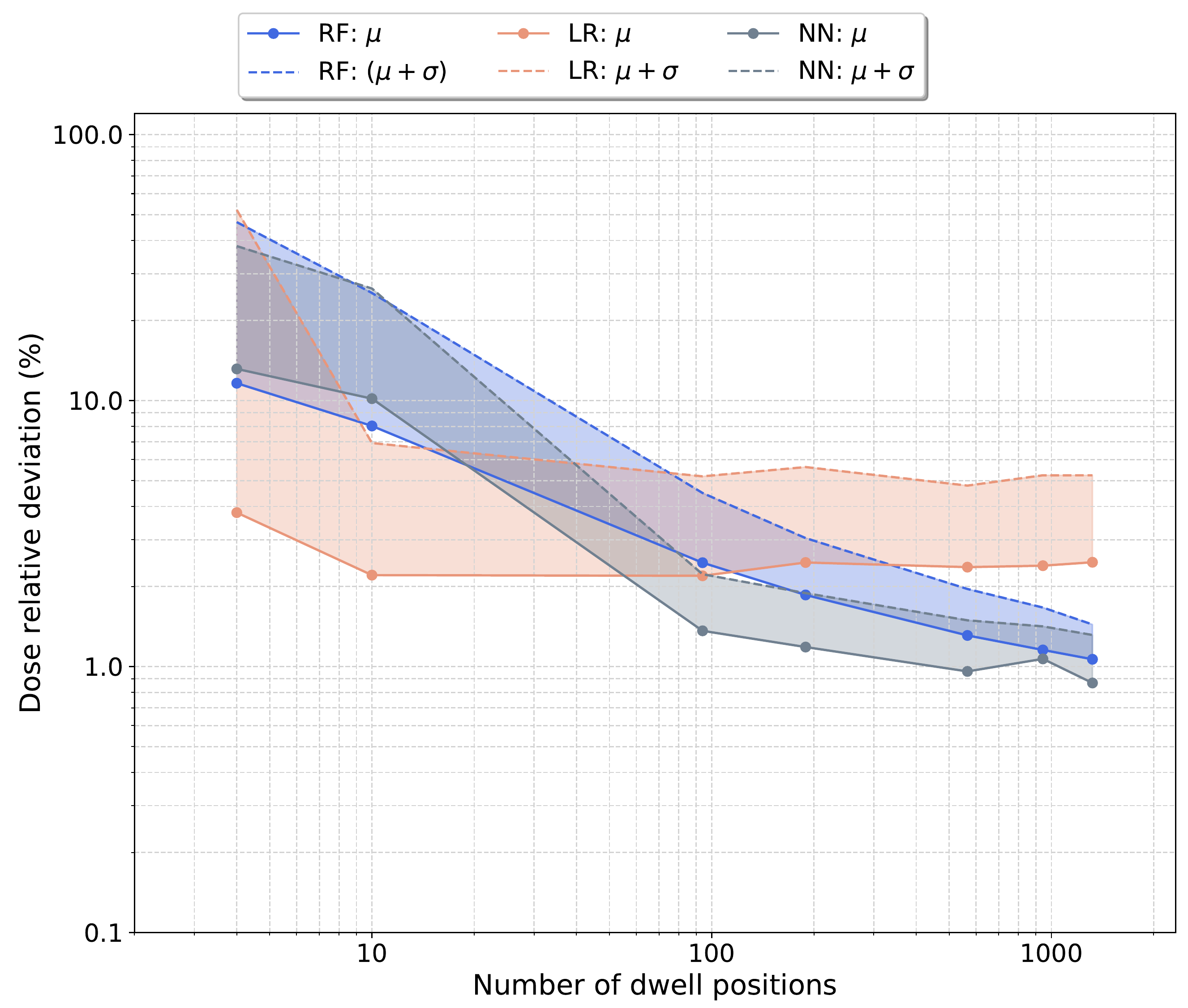}
    \end{tabular}
    \caption{\label{R_influence_sample} Influence of the training sample size on the dose prediction model.}
    \end{figure}

    Figure \ref{R_influence_sample} shows the influence of the training sample size on the models’ performance. The solid lines are the relative median deviations from the expected TG-43 U1 dose \cite{TG-43-Update} obtained with the random forest, linear regression, and ANN algorithms. The shaded regions represent the ranges of deviations between the median deviations and the median deviations plus the standard deviations. We created Figure \ref{R_influence_sample} based on the deviations obtained at all distances to the source for study case 2 in Table \ref{table_MM_ML_parameters}. The relative deviations for training done with fewer than 100 dwell positions resulted from calculations done 150 times each.
    
\subsubsection{Overall performance}

    Table \ref{R_table_ML_method_comparisons} summarizes the performance of the different models used for mPSD dose prediction. We rated the models as advantageous (++), good (+), or inconvenient (-). To evaluate each model's performance, we took into account different criteria. First, we considered the dose measurement accuracy, which we evaluated using the expected TG-43 U1 dose \cite{TG-43-Update} as a reference. Additionally, we considered the implementation complexity of each model to provide the reader with an overall point of comparison (interpretability of the trained model, degrees of freedom, training process, and sensitivity to metaparameters adjustment). Table \ref{R_table_ML_method_comparisons} also shows each model's performance for a small training set with the intent of evaluating the extent of the required user's expertise in mPSD manipulation to obtain accurate results with each model analyzed.
    
    \begin{table}[ht]
    \caption{\label{R_table_ML_method_comparisons} Overall performance of the models used for mPSD calibration and dose measurement.}
    \setlength{\tabcolsep}{7pt}
    \centering
    \vspace{0.2cm}
    	\begin{tabular}{ccccccccc}
    	\hline
    	\hline
    	                  \textbf{Criteria}  && \textbf{Applicability} && \textbf{Linear}     && \textbf{Random} && \textbf{Neural} \\
    	                                    &&                    && \textbf{regression}     && \textbf{forest} && \textbf{network} \\
    	\hline
    	\multirow{2}{*}{\textbf{Accuracy}}  && \textit{inside region}  && + && ++ && + \\
    	                                    && \textit{outside region} && +  &&  - && - \\
    	\hline
    	\textbf{Method implementation}      && \multirow{2}{*}{-} && \multirow{2}{*}{++} && \multirow{2}{*}{+} && \multirow{2}{*}{-}\\
    	\textbf{complexity}                 \\
    	\hline
    	\textbf{Performance for small}      && \multirow{2}{*}{-} && \multirow{2}{*}{++} && \multirow{2}{*}{-} && \multirow{2}{*}{-}\\
    	\textbf{training set}                 \\
    	\hline
    	\textbf{mPSD's manipulation: required} && \multirow{2}{*}{-} && \multirow{2}{*}{-} && \multirow{2}{*}{++} && \multirow{2}{*}{++}\\
    	\textbf{user expertise}                 \\
    	\hline
    	\hline
    	\multicolumn{9}{c}{\small{++, advantageous; +, good; -, inconvenient.}}\\
    	\end{tabular}
    \end{table}

\section{Discussion}

    In this study, we evaluated the feasibility of ML algorithms for mPSD calibration in HDR brachytherapy. The overall models' deviations from the expected doses generally remained below 20 \%. In agreement with the observations previously \cite{Linares-2019}, the dose prediction accuracy and measurement uncertainty were dependent on the calibration conditions and dose prediction model. We observed the best accuracy for calibrations done in study case 1E (Table \ref{table_MM_ML_parameters}), in which we used the measured data covering the whole range of distances explored to extract the calibration factor for signal translation into the dose. However, when considering only a portion of the data (i.e., study cases 1A-D), we found the best agreement in study case 1B. In this case, the linear regression model resulted in median dose deviations smaller than 7 \%. The study done by Linares Rosales et \textit{al.} \cite{Linares-2019} also explored the dose agreement with TG-43 U1 \cite{TG-43-Update} over distances to the source ranging from 5 to 65 mm. At 65 mm from the source, they observed relative deviations of 21.3 \%, 12.4 \%, and 13.7 \% for the BCF-10, BCF-12, and BCF-60 scintillators, respectively.  In the present study, the dose agreement with the expected TG-43 U1 dose was superior to the findings in \cite{Linares-2019} due to optimal selection of the calibration conditions.  On the other hand, the selection of data for study case 1C or 1D for calibration purposes led to increased deviation at short distances to the source. This effect was most pronounced while using region 1C, where deviations could reach 11 \% relative to the TG-43 U1 reference dose. Based on the findings shown in figure \ref{R_influence_calibration}, we recommend the use of data at distances from the source ranging from 25 to 75 mm (region 1B) for model training and mPSD calibration. When using this region, the relative dose deviation obtained with the three ML algorithms remained small (0.2-4.2 \%) at all of the explored distances from the source. We do not recommend using data in the source's high-gradient field for mPSD calibration because it will lead to unbalanced fitting of the dose. As a result, increased deviations at large distances from the source can be expected.
    
    The number of available measurements for training purposes influences random forest and ANN models the most. As shown in \ref{R_influence_sample}, the median ($\pm$ standard deviation) dose deviation for the predictions with linear regression remained almost constant at about 2.12 \% $\pm$ 3.98 \% when we used at least 10 dwell positions. In principle, the linear regression model can be trained with only four dwell positions. However, the dose prediction accuracy would be highly influenced by the user's experience with mPSD technology, which could lead to poor calibration conditions. The random data selection used to create Figure \ref{R_influence_sample} highlights this effect. The use of four dwell positions in the model's training could lead to dose deviations on the order of 50 \% from the expected TG-43 U1 dose \cite{TG-43-Update}. The results obtained with the random forest and ANN algorithms were similar. The accuracy obtained with those models begins to converge toward deviation values close to 1 \% from a number of dwell positions greater than 100, which may not be feasible for clinical end users.
    
    Table \ref{R_table_ML_method_comparisons} summarizes the findings of this study. In general, the dose predictions with all three models were accurate within 7 \% relative to the dose predicted by TG-43 U1  \cite{TG-43-Update}, when measurements are done over the same range of distances used for calibration (the \textit{inside region} criterion in table \ref{R_table_ML_method_comparisons}). In such cases, the predictions with the random forest algorithm exhibit minimal deviations, but the performance is compromised for predictions made outside the calibration region. Signal spikes and abrupt signal changes in the measured dose are frequently observed with the random forest and ANN algorithms under conditions different from the calibration ones as observed in regions 1A, 1C, and 1D in Figure \ref{R_influence_calibration}. The best accuracy for both inside and outside the calibration is obtained with the linear regression model. The behaviour of the deviations along the distances to the source is smoother for linear regression than for the other two algorithms even if larger deviations are observed. This behavior reflects the well-known fact that linear regression algorithms can extrapolate data but the random forest algorithm cannot and that linear regression algorithms outperform the random forest algorithm in situations with low signal-to-noise ratios (typical of long distances to the source). Indeed, because the random forest algorithm is not a differentiable function but rather a combination of decision trees with discrete output values (in agreement with given output values in the training phase), it cannot accurately predict the dose beyond the range of the calibration region. The accuracy of the dose predictions attained with the ANN algorithm is similar to that with the random forest algorithm. A non optimal ANN architecture may explain this effect. Indeed, the ANN's internal operations, which consist of simple functions (e.g., the linear function), try to reproduce the physical system behavior by adjusting its internal weights. The difficulty when training such an algorithm is finding the optimal network architecture that will enable it to reproduce the behavior of the underlying physical system without overfitting of the dose.
    
    Apart from the mentioned above, in selecting a proper method for mPSD dose measurement, the method implementation complexity must be considered. Interpreting the results obtained with the linear regression model is easier than that for the ANN or random forest algorithm. These kinds of methods require fewer data than more refined models to produce acceptable results. However, more refined models, such as the random forest and ANN models, may have better performance when appropriately trained. Indeed, Figure \ref{R_influence_calibration} shows that the random forest is usually the best algorithm when the calibration region encompasses the prediction region, even if the data are randomly selected during the training process (e.g., region 1E in Figure \ref{R_influence_calibration}). The last two criteria we considered in Table \ref{R_table_ML_method_comparisons} were the models’ performance using a small training data set and the user expertise required to manipulate the mPSD.
    
    Although the linear regression model's median dose deviation remains relatively constant across the training sample sizes, the achieved agreement with the expected TG-43 U1 dose depends on the dwell position selected for training and calibration among other parameters. The effect of the data selections becomes critical when using four dwell positions, for instance. This is why we considered the linear regression model to be inconvenient according to the user expertise criterion. When using four dwell positions for training purposes, significant deviations are also observed with the random forest and ANN models (performance ranked as inconvenient for a small training data set) because they can incorrectly interpolate positions in regions not present in the training data set. However, as shown in Figure \ref{R_influence_sample}, at a certain number of training dwell positions, the accuracy achieved with random forest and ANN models is superior to that achieved with the linear regression model while being independent from the user expertise in selecting the appropriate dwell position.
    
    Given the results described herein, taking a large number of dose measurements in a clinical environment may be considered time-consuming. Nevertheless, for example, setting 200 dwell positions for training purposes with dwell times of 1 s each represents about 3.33 min of irradiation. The time-consuming factor associated with this process is setup of the experiments, meaning mPSD and catheter positioning to ensure TG-43 U1 conditions, connecting transfer tubes and double-checking verification to ensure lack of a catheter swap, and correcting catheter shifts. This process takes several minutes and must be optimized to reduce the time to it takes.

\section*{Conclusions}

    In this study, we evaluated the benefits and limitations of using linear regression, random forest, and ANN algorithms for mPSD calibration in HDR brachytherapy. The drawback of these methods is the need to calibrate the detector at many locations relative to the source, which may result in a relatively long calibration time. The main benefits of using the models described herein are that calibration must only be done once for each detector and that the pretrained models' files can be distributed for mPSD implementation in brachytherapy clinics. Finally, our results suggest that the random forest algorithm has the best performance of the three algorithms tested when the data from the training region cover the prediction region. We also recommend using training data with the measurements in at least 100 dwell positions. In such cases, agreement 7 \% with the expected TG-43 U1 dose is ensured with all the algorithms, and selection of the calibration dwell positions becomes less dependent on the user’s expertise. 
  
\section*{Acknowledgements}

    This work was supported by the National Sciences and Engineering Research Council of Canada via National Sciences and Engineering Research Council of Canada-Elekta Industrial Research Chairs grants 484144-15 and RGPIN-2019-05038 and by Canadian Foundation for Innovation John R. Evans Leaders Fund grant 35633. HMLR further acknowledges support from Fonds de Recherche du Quebec - Nature et Technologies and by CREATE Medical Physics Research Training Network grant 432290 from the Natural Sciences and Engineering Research Council of Canada. We also thank Donald R. Norwood from Scientific Publications Services in the Research Medical Library at The University of Texas MD Anderson Cancer Center for editing our manuscript.

\end{document}